\documentstyle[preprint,aps,epsf,array]{revtex}
\def\baselinestretch{1.5}

\newcommand{\be}[1]{\begin{equation} \label{(#1)}}
\newcommand{\ee}{\end{equation}}
\newcommand{\ba}[1]{\begin{eqnarray} \label{(#1)}}
\newcommand{\ea}{\end{eqnarray}}
\newcommand{\nn}{\nonumber}

\def\lsim{\mathrel{\vcenter{\hbox{$<$}\nointerlineskip\hbox{$\sim$}}}}
\def\gsim{\mathrel{\vcenter{\hbox{$>$}\nointerlineskip\hbox{$\sim$}}}}

\begin{document}
\preprint{\vbox{\hbox{VAND--TH--01--2}
\hbox{February 26, 2001}
}}
\title{Muon anomalous magnetic moment in string inspired 
extended family models}
\author{T.W. Kephart
\footnote[1]{E-mail: kephartt@ctrvax.vanderbilt.edu} 
and
H. P\"as
\footnote[2]{E-mail: heinrich.paes@vanderbilt.edu} 
}
\address{Department of Physics and Astronomy\\ Vanderbilt University\\
Nashville, TN 37235, USA}

\maketitle

\begin{abstract}
We propose a standard model minimal extension with two lepton weak
$SU(2)$ doublets and a scalar singlet to
explain the deviation of the measured anomalous magnetic moment of the muon
from the standard model expectation. This scheme can be naturally motivated
in string inspired models such as $E_{6}$ and AdS/CFT.
\end{abstract} 

\newpage
The recent result of the $g-2$ Collaboration at BNL may be the overdue
first direct signal of physics beyond the standard model (SM). The data seem
to indicate a 2.6 $\sigma$ deviation from the theoretical expectation 
of the SM \cite{data},
\be{}
a_{\mu}^{exp}-a_{\mu}^{SM}=426 \pm 165 \cdot 10^{-11}.
\ee
Several logical possibilities have been considered \cite{marc}
to explain the effect,
such as supersymmetry \cite{susy}, leptoquarks
\cite{lqs}, muon substructure \cite{subs}, 
technicolor \cite{tc}, large extra dimensions \cite{ed},
new interactions 
and fermions \cite{nif}; for a review see \cite{marc}. 
In this letter we propose a minimal extension of the SM, which to date
has not been discussed in this context. 

It is sufficient to extend the second
fermion
family (extending the other families is optional but aesthetically
appealing)
by adding an extra pair of lepton doublets and a scalar singlet:

\be{}
\left(
\begin{array}{l}
M \\
N
\end{array}
\right) _{L}+\left(
\begin{array}{l}
M \\
N
\end{array}
\right) _{R},~~ S^0
\ee
with no other states or extension of the gauge group necessary.
There are several ways to motivate this choice and we will focus on two.

First if the standard model is embedded in an $E_{6}$ model,
then the fermion families contain extra states. For the
symmetry
breaking chain
$E_{6}$ $\rightarrow SO(10)\rightarrow SU(5)\rightarrow SU(3)\times
SU(2)%
\times U(1)$,
an  $E_{6}$ family of fermions decomposes as:

\ba{}
27\rightarrow 16+10+1\rightarrow
(\bar{5}+10+1)+(5+\bar{5})+1 \nn \\
\rightarrow {\rm(SM~family)} 
+(3,1)+(\bar{3},1)+(1,2)+(1,\bar{2})+2(1,1).
\ea

The conjugate pair of doublets are just those required above. The extra
singlet quarks and SM singlets can be made relatively heavy by proper
choice
of vacuum expectation values (VEVs) and coupling constants.
Extra gauge bosons can also be sufficiently heavy that they have 
decoupled from the subsequent analysis.
For other discussions of $a_{\mu}$ in the $E_{6}$ model see \cite{e6,rizzo}.

The second possibility is a non-SUSY model with extra lepton doublets
based
on either an abelian or a nonabelian orbifold AdS/CFT type IIB string
theory \cite{Kachru:1998ys,Frampton:1999wz,Frampton:2000zy,Frampton:2000mq}.
These models have gauge groups that are the product of
$SU(d_{i}N)$
groups where the $d_{i}$ are the dimensions of the irreps of the
orbifolding
group $\Gamma $ and $N$ can be chosen. Thus the gauge group $G$ can be
of the form $G=SU(3)\times SU(3)\times SU(3)\times G^{\prime }$
(where
$G^{\prime }$ can be either broken or ignored for our purposes).
All
the fermions reside in bifundamental representations of the gauge group
and
so there can be $N_F$ $SU^3(3)$ families plus other states:

\be{}
N_{F}[(3,\bar{3},1)+(1,3,\bar{3})+(\bar{3},1,3)]+...
\ee

Note that these contain the same fermions one acquires
from breaking  $E_{6}$ to
its
maximal subgroup $SU^{3}(3)$ so we are again naturally led to
\be{}
N_{F}[{\rm (SM~family)}+(3,1)+(\bar{3},1)+(1,2)+(1,\bar{2})+2(1,1)]+....
\ee
which provides the necessary pair of extra lepton doublets.
A typical length scale for orbifold AdS/CFT models is a few TeV, so
fermions in this mass region or somewhat lighter fit easily into the
scheme proposed here.

Scalar singlets also arise in both schemes.

\begin{figure}
\epsfxsize=120mm
\epsfbox{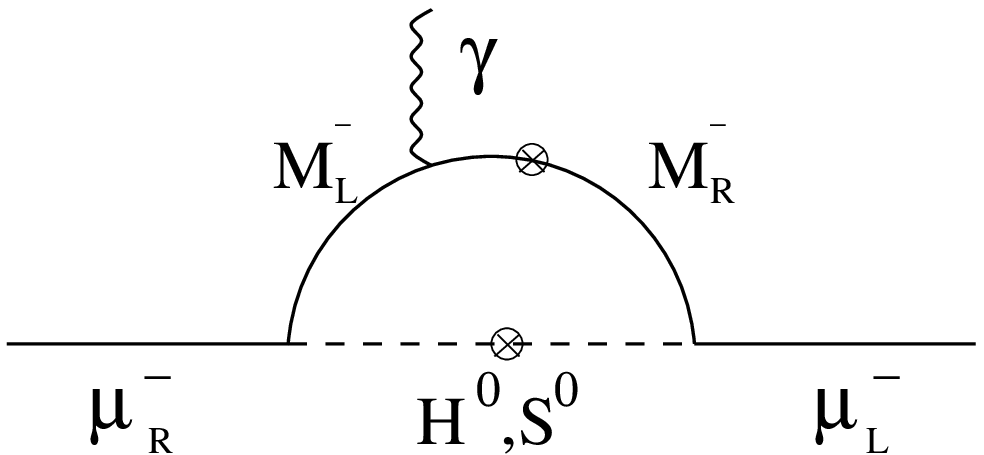}
\vspace*{1cm}
\caption{Feynman graph contributing to the muon anomalous magnetic moment
in the extended family model.
\label{fd}}
\end{figure}

The contribution of the new muon type heavy charged lepton $M^-$ to $a_{\mu}$
is due to the Feynman diagram fig. \ref{fd}. Similar diagrams with SM gauge 
boson instead of Higgs boson/scalar singlet exchange have been considered, but 
cannot provide a correction to $a_{\mu}$ to allow agreement with experiment
\cite{rizzo}. The FCNC-contribution 
with $\tau$ (instead of $M^-$) exchange is forbidden in the standard model, 
since the mass matrix is directly proportional to the Yukawa coupling 
matrix. The viable extended model with a non-SM Higgs doublet has been 
discussed in \cite{sher}. In our case the proportionality is broken by 
assigning a bare mass to $M^-$. 

The contribution to $a_\mu\equiv
{g_\mu-2\over 2}$ is given by \cite{lev}
\be{}
a_{\mu}={\xi_{\mu M}\over 16\pi^2}\int_0^1
{x^2(1-x) +  x^2{m_{M} \over m_\mu} \over
x(x-1)+x{m_{M}^2\over m_\mu^2}+(1-x){m^2_{1}\over m^2_\mu}}\ dx.
\ee
Here $\xi_{\mu M}$ refers to the product 
\be{}
\xi_{\mu M}=h_{\mu M} h'_{\mu M} O_{H^{0}1} O_{S^{0}1},
\label{xi}
\ee
where $h_{\mu M}$, $h'_{\mu M}$ are the couplings to the real fields 
$H^0$ and $S^0$, respectively and $O_{H^{0}1} O_{S^{0}1}$ denotes the product
of orthogonal doublet-singlet mixing matrix elements with the light scalar
eigenstate ($m_{1} \ll m_{2}$ has been assumed).  

\begin{figure}
\epsfxsize=120mm
\epsfbox{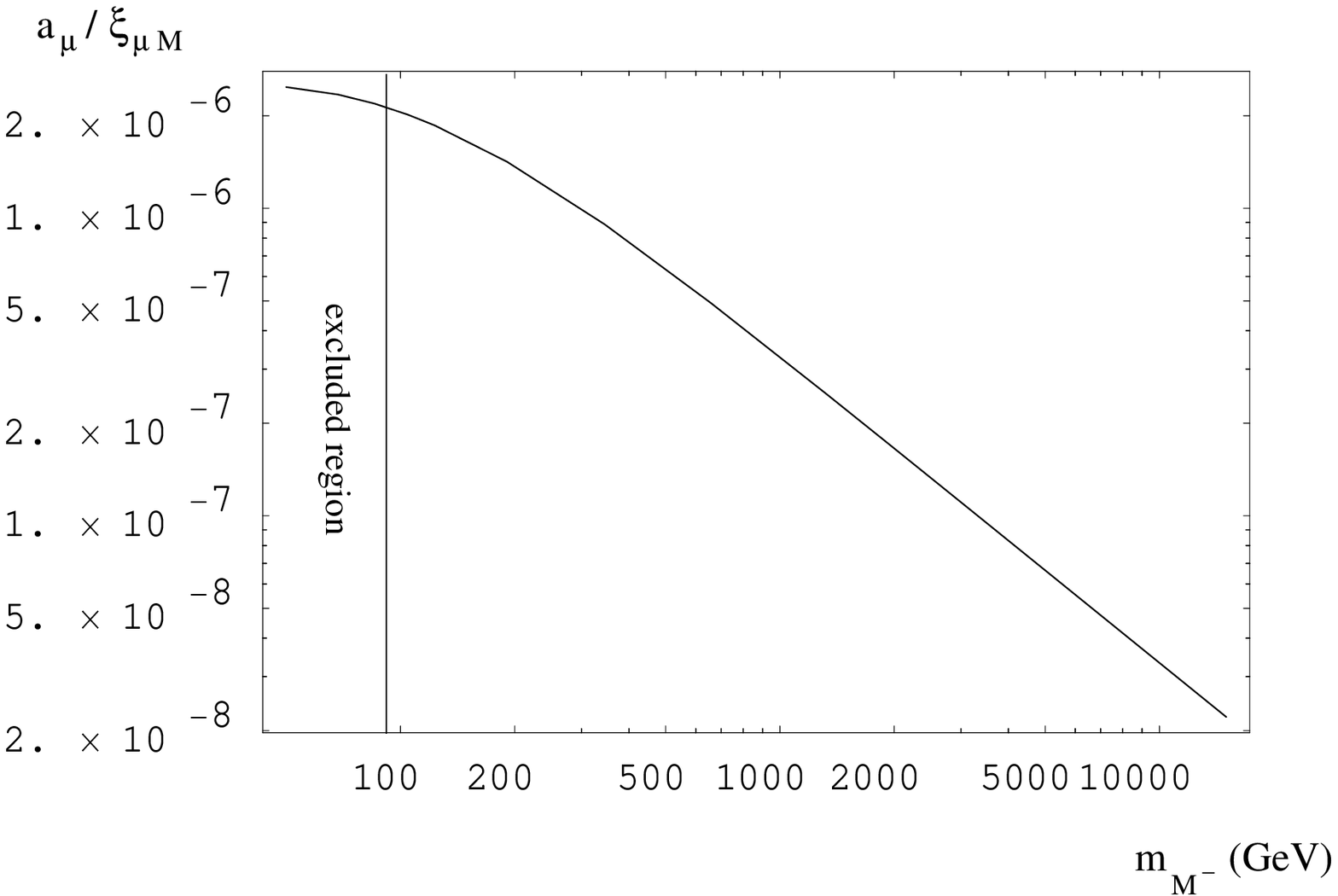}
\vspace*{1cm}
\caption{Muon magnetic moment $a_{\mu}$ in terms of 
$\xi_{\mu M}$ as a function of the new muon type heavy charged lepton mass
$M^{-}$.
\label{f2}}
\end{figure}

In fig. \ref{f2}, $a_{\mu}/\xi_{\mu M}$ is plotted as a function of the
heavy charged lepton mass $m_{M}$. The plot shows the entire perturbative
region, i.e. $\xi_{\mu M}\lsim 0.2$, so that $m_{M}\lsim 15$~TeV.
With $\xi_{\mu M}\gsim 0.01-0.1$, assuming a physical
Higgs mass (light scalar mass eigenstate) of $m_{1}=115$~GeV 
and $m_{M}$ of order 1-10 TeV, the anomaly 
can be explained. (The result is insensitive to the Higgs mass when 
$m_{M} \gg m_1$.)
This is well above the recent limits of heavy charged leptons, 
$m_{M}\gsim$~95 GeV \cite{pdg}, but lies in the range to be explored at
future accelerators such as the LHC \cite{lhc}.

Since the $\mu^--M^--H^0$ coupling $h_{\mu M}$ induces offdiagonal terms
in the muon mass matrix, the model may have interesting implications for weak 
universality, which will be discussed in the following.
The masses of the charged leptons are obtained by diagonalizing the mass term
\ba{}
\overline{\left(\mu^C_R, \mu_L, M^C_R, M_L \right)} {\cal M}
\left(
\begin{array}{c}
\mu_R  \\
\mu_L^C\\
M_R    \\
M_L^C  \\
\end{array}
\right)
=
\overline{\left(\mu^C_R, \mu_L, M^C_R, M_L \right)}
\left(
\begin{array}{cccc}
0   & g v & 0    & h v \\
g v &  0  & h'v' &  0  \\
0   & h'v'& 0    &  M  \\
h v &  0  & M    &  0  \\
\end{array}
\right)
\left(
\begin{array}{c}
\mu_R  \\
\mu_L^C\\
M_R    \\
M_L^C  \\
\end{array}
\right)
\ea
via 
\ba{}
V^{\dagger} {\cal M} V= diag(m_{\mu},m_{\mu},m_{M},m_{M}). 
\ea
Here $v, v'$ are the VEVs of the Higgs doublet and singlet, respectively,
and $g$ is the muon Yukawa coupling. A stringent bound 
on the mixing matrix elements $V_{11}$, $V_{12}$
determining the admixture of the heavy state 
in the muon flavor eigenstate is
obtained from its contribution to the pion decay.
The agreement of the measured pion decay rate ratio,
\be{}
R_\pi = {\Gamma(\pi^+ \to e^+ \nu) \over \Gamma(\pi^+ \to \mu^+ \nu)} 
= (1.235 \pm 0.004) \times 10^{-4},
\ee
with the SM value $R_\pi(SM) = (1.230 \pm 0.008) \times 10^{-4}$ 
implies $G_N^\mu \lsim 4 \times 10^{-3} G_F$ for any new 
physics contribution $G_N$, i.e. $\sqrt{V_{11}^2+V^2_{12}}<4 \times 10^{-3}$
(see, e.g. \cite{kolb}). This constrains the offdiagonal entry $hv$ to be 
small, while the product of coupling constants and mixing matrices
$\xi_{\mu M}$ in eq. (\ref{xi})
has to be large to yield the correct value for the anomalous 
magnetic muon moment.

Eigenvalues $m_{\mu} \simeq g v =100$~MeV and $m_{M}=1-10$~TeV 
as well as the 
right anomalous contribution to the muon magnetic moment  
are obtained by assigning e.g. $g v \simeq 100$~MeV, $h v \simeq 3-30$~GeV, 
$v'<vh/h'$, $h' \simeq 2$ and $M \simeq m_{M} =1-10$ TeV and maximal
scalar mixing $O_{H_0 1} O_{S^0 1} \simeq 0.5$. 
The mixing matrix elements are
$V_{11} \simeq V_{12} \simeq 2 \cdot 10^{-3}$
and $V_{21} \simeq V_{22} \ll 10^{-3}$. They are close to but
compatible with the bounds
obtained from the pion decay rate ratio.

The one-loop correction to the muon mass is obtained by removing the
photon line from the diagram fig. 1, and can be estimated to be 
\cite{marc}
\be{}
m^{loop}_{\mu} \simeq \frac{1}{16 \pi^2} \cdot \xi_{\mu M}\cdot m_{M} \cdot 
\left(\frac{m_{M}^2 \cdot ln~ m_{M}^2 - m_1^2 \cdot ln~ m_1^2}
{m_{M}^2 - m_1^2}
-\frac{m_{M}^2 \cdot ln~ m_{M}^2 - m_2^2 \cdot ln~ m_2^2}{m_{M}^2 - m_2^2}\right),
\ee
which can be small as long as $m_2 \ll m_M$, i.e. $m_2 \lsim 300-500$~GeV
for $m_M \simeq 1-10$~TeV.

To conclude, we have shown an extended muon family is sufficient to
provide the contribution needed to bring theory back in line with the
present $g-2$ data for the muon. The proposed model is motivated from more
fundamental theory and the results are simply obtained. 
The new particles required by the model
are potentially within the reach of the LHC and large violations of weak 
universality are predicted.

\section*{Acknowledgements}
We would like to thank L. Clavelli, M. Sher and T.J. Weiler 
for useful discussions and correspondence.
This work was supported in part by the
DOE grant no.\ DE-FG05-85ER40226.

\newpage

\newpage

\begin{thebibliography}{200}

\bibitem{data}
H.N. Brown, et al, Muon g-2 Collaboration,
hep-ex/0102017

\bibitem{marc} 
A. Czarnecki, W. J. Marciano, hep-ph/0102122 and hep-ph/0010194

\bibitem{susy}
L. Everett, G.L. Kane, S. Rigolin, and L.-T. Wang, hep-ph/0102145;  
J.L. Feng, K.T. Matchev, hep-ph/0102146; 
E.A. Baltz, P. Gondolo, hep-ph/0102147;  
U. Chattopadhyay, P. Nath, hep-ph/0102157;
S. Komine, T. Moroi, M. Yamaguchi, hep-ph/0102204.


\bibitem{lqs}
U. Mahanta, hep-ph/0102176;  
D. Chakraverty, D. Choudhury, A. Datta,  hep-ph/0102180.

\bibitem{subs}
K. Lane,  hep-ph/0102131

\bibitem{tc}
Z. Xiong, J. M. Yang, hep-ph/0102259.

\bibitem{ed}
R. Casadio, A. Gruppuso, G. Venturi,
Phys. Lett. {\bf B 495} (2000) 378

\bibitem{nif}
T. Huang, Z.-H. Lin, L.-Y. Shan, X. Zhang, hep-ph/0102193;  
D. Choudhury, B. Mukhopadhyaya, S. Rakshit, hep-ph/0102199;
U. Mahanta, hep-ph/0102211;
E. Ma, M. Raidal, hep-ph/0102255.


\bibitem{e6}
J.A. Grifols, A. Mendez, J. Sola, Phys. Rev. Lett. {\bf 57}, 2348 (1986);
D.A. Morris, Phys. Rev. {\bf D 37}, 2012 (1988).



\bibitem{rizzo} 
T.G. Rizzo, Phys. Rev. {\bf D 33}, 3329 (1986) 

\bibitem{Kachru:1998ys}
S.~Kachru and E.~Silverstein,
Phys.\ Rev.\ Lett.\ {\bf 80}, 4855 (1998)
[hep-th/9802183].

\bibitem{Frampton:1999wz}
P.~H.~Frampton,
Phys.\ Rev.\ D {\bf 60}, 121901 (1999)
[hep-th/9907051].

\bibitem{Frampton:2000zy}
P.~H.~Frampton and T.~W.~Kephart,
Phys.\ Lett.\ B {\bf 485}, 403 (2000)
[hep-th/9912028].

\bibitem{Frampton:2000mq}
P.~H.~Frampton and T.~W.~Kephart,
hep-th/0011186.

\bibitem{sher}
S. Nie, M. Sher, Phys. Rev. {\bf D 58}, 097701 (1998) 

\bibitem{lev}
J.P. Leveille, Phys. Rev. {\bf D 137}, 63-76 (1978)

\bibitem{pdg}
Particle Data Group, Eur. Phys. J. {\bf C 15} 1, (2000) 

\bibitem{lhc}
P.H. Frampton, P.Q. Hung, M. Sher, Phys. Rept. {\bf 330} 263, (2000) 

\bibitem{kolb}
St. Kolb, M. Hirsch, H.V. Klapdor--Kleingrothaus, 
Phys. Rev. D 56 (1997) 4161 

\end{thebibliography}
\end{document}